\documentclass[namedreferences]{SolarPhysics}
\pdfoutput=1
\usepackage[optionalrh]{spr-sola-addons} 

\usepackage{solaheader}  
\newcommand{\solareceiveddate}{19 Oct 2010} 
\newcommand{\solaaccepteddate}{16 Nov 2010}

\usepackage{graphicx}        
\usepackage{amssymb}         
\usepackage[usenames,dvipsnames]{color}          
\usepackage{url}             

\DeclareFontFamily{OT1}{pzc}{}
\DeclareFontShape{OT1}{pzc}{m}{it}%
             {<-> s * [1.1500] pzcmi7t}{}
\DeclareMathAlphabet{\mathscr}{OT1}{pzc}%
                                 {m}{it}

\usepackage{epstopdf}
\usepackage{amssymb,amsmath}

\newcommand{\half}{{\textstyle\frac{1}{2}}}

\newcommand{\re}{\mathop{\rm Re}\nolimits}

\newcommand{\rd}{\mathrm{d}}

\newcommand{\pderivd}[2]{\frac{\partial^2#1}{\partial#2^2}}

\newcommand{\deriv}[2]{\frac{\rd#1}{\rd#2}}

\newcommand{\vdot}{{\boldsymbol{\cdot}}}

\newcommand{\grad}{\mbox{\boldmath$\nabla$}}
\newcommand{\bxi}{\mbox{\boldmath$\xi$}}

\newcommand{\thth}{\hspace{1.5pt}}

\newcommand\Div{\grad\vdot\thth}

\newcommand{\kperp}{k_{\scriptscriptstyle\!\perp}}

\newcommand{\ri}{\mathrm{i}}

\newcommand{\solphys}{\textit{Solar Phys.}}

\allowdisplaybreaks[1]

\begin{document}
\begin{opening}

\title{Resonant Absorption as Mode Conversion? II. Temporal Ray Bundle}

\author{Chris S.~\surname{Hanson}\sep
Paul S.~\surname{Cally}}

\institute{Centre for Stellar and Planetary Astrophysics, School of
Mathematical Sciences, Monash University, Victoria 3800, Australia \email{paul.cally@monash.edu} 
}

\runningauthor{C.S. Hanson and P.S. Cally}

\runningtitle{Resonant Absorption as Mode Conversion? II}

\begin{abstract}
A fast-wave pulse in a simple, cold, inhomogeneous MHD model plasma is constructed by Fourier superposition over frequency of harmonic waves that are singular at their respective Alfv\'en resonances. The pulse partially reflects before reaching the resonance layer, but also partially tunnels through to it to mode convert to an Alfv\'en wave. The exact absorption/conversion coefficient for the pulse is shown to be given precisely by a function of transverse wavenumber tabulated in Paper I of this sequence, and to be independent of frequency and pulse width. 
\end{abstract}



\end{opening}


\section{Introduction}
Recently, \inlinecite{ca10} (Paper I) explored mode conversion and resonant absorption in a simple model consisting of a uniform magnetic field oriented in the $z$-direction threading a cold magneto\-hydro\-dynamic (MHD) plasma with a monotonically decreasing density stratification in the $x$-direction. The Alfv\'en speed $a(x)$ is consequently monotonically increasing in $x$. A monochromatic, plane, fast wave with frequency $\omega$ and $y$ and $z$ wavenumbers $k_y$ and $k_z$ incident from $x\to-\infty$ classically reflects where $\omega^2=a^2(k_y^2+k_z^2)$, but in fact partially tunnels through to the Alfv\'en resonance $\omega^2=a^2k_z^2$ located distance $X$ to the right of the reflection point. We adopt the simple exponential Alfv\'en profile $a=a_0 \mathrm{e}^{x/2h}$, which may be regarded as deriving from a local tangent approximation to $\ln a$ in the neighbourhood of the resonance. It was shown for this profile in Paper I that the fraction $\mathscr{A}$ of the incident wave-energy flux that is absorbed by the resonance and converted into an Alfv\'en wave depends solely on the two parameters $\theta=\tan^{-1}(k_y/k_z)$ and $\sigma=(\kperp h)^{2/3}\sin^2\theta$, where $\kperp=\sqrt{k_y^2+k_z^2}$. The remainder of the energy is reflected back to the left, also in the form of a fast MHD wave. Without loss of generality, the density scale distance $h$ is set to unity throughout.

\begin{figure}[tbp]
\begin{center}
\includegraphics[width=.49\textwidth]{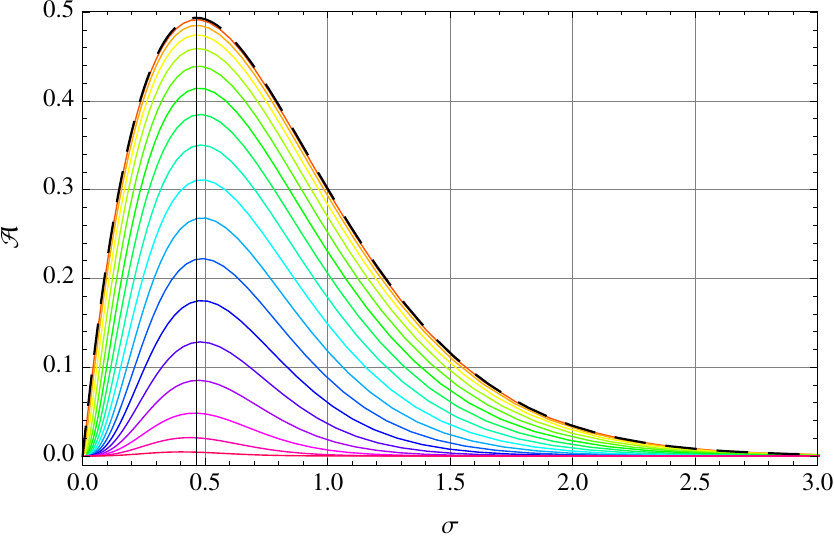}
\caption{Absorption coefficient $\mathscr{A}(\sigma,\theta)$ for $a^2\propto \mathrm{e}^{x/h}$ and $\theta=5^\circ$, $10^\circ$, \ldots, $85^\circ$ (coloured curves, top to bottom). The heavy dashed curve corresponds to $\theta=0^\circ$. The heavy vertical line is at $\sigma=\sigma_0=0.464434$, where $\mathscr{A}(\sigma,0)$ attains the maximum value $A_0=0.493698$. $\mathscr{A}$ is tabulated in Appendix C of Paper I. [This is a reprinting of Figure 2a of Paper I.]}
\label{fig:Asig}
\end{center}
\end{figure}

Intriguingly, absorption may not exceed $A_0=49.37\%$ (see Figure \ref{fig:Asig}), the maximum corresponding to $\sigma=\sigma_0=0.464434$ and $\theta\to0^\circ$, and falling rapidly as $\sigma$ increases or decreases from this value. Of course, $\kperp\to\infty$ in the limit where $\theta$ vanishes but $\sigma$ remains finite. As is well known, the Alfv\'en resonance corresponds to a singular point of the governing differential equation, where the wave-energy density is not only unbounded, but also not integrable. This is because energy is assumed to have been absorbed at constant rate over an infinite time. 

The existence of the resonance is a consequence of the magnetic field being perpendicular to the inhomogeneity direction $x$ (assuming specified Fourier dependencies in the $y$ and $z$ directions). We choose this special case precisely because we wish to explore the behaviour of a wave pulse meeting a resonance in the simplest possible model. With oblique field, mode conversion still operates and may exceed $A_0$. In the solar context, the case we are considering is most relevant to horizontal magnetic field in a vertically stratified atmosphere.

To better explore the nature of the resonance, and the relationship between resonant absorption and mode conversion, \inlinecite{ca10} constructed a well-behaved spatial ray bundle from the singular wave solutions by Fourier composition. The modes from which this bundle was synthesized all had the same $\omega$ and $k_y$ but were distributed in $k_z$ as a gaussian about a central $k_{z0}$. This had the dual effects of both localizing the conversion/absorption region in $z$ and smearing the resonance. A very clear mode conversion process became apparent graphically, with the conversion coefficient being an appropriate average of the $\mathscr{A}$ values drawn from Figure \ref{fig:Asig}. 

On presenting these results in seminars at several institutes, one of the authors (PSC) was consistently asked about how this would generalize to \emph{temporal} wave packets; \emph{i.e.}, a sharp (or otherwise) incident wave burst
with single $k_y$ and $k_z$. This is relevant to waves excited by discrete events in the solar corona for example. The answer is quite simple, and in the interests of completeness is presented here. For conciseness, necessary results from \inlinecite{ca10} will be assumed. Further discussion of context and of the related literature may also be found in Paper I.

Whereas the spatial ray bundle case deposits energy continuously into the resonant layer in the vicinity of a localized height [$z_0$] and mode conversion can be identified as an Alfv\'enic-wave energy flux on $z\gtrsim z_0$ taking energy away from this site, the temporal ray packet instead deposits energy equally at all $z$ along the resonance layer but only for a short time. The issue is then the total energy per unit $y$--$z$ area left on the resonance once the driving fast wave has departed, as a fraction of the integrated incident flux. We ask how that fraction depends on frequency, wavenumbers, and packet duration, and in particular if it can in any way exceed $A_0=49.37\%$.

\section{Temporal Wave Packet}
Let $\hat\chi(x,\omega)$ be the (singular) monochromatic solution for the dilatation $\chi=\Div\bxi$ corresponding to frequency $\omega$, where $\bxi$ is the plasma displacement. Assuming a $\chi=\hat\chi\exp[\ri(k_yy+k_zz-\omega t)]$ transverse and time dependence, the governing equation is (\emph{c.f.}, Equation (3) of Paper I)
\begin{equation}
\deriv{\ }{x}\left[\frac{1}{\omega^2/a^2k_z^2-1}\deriv{\hat\chi}{x}\right]+\left(k_z^2-\frac{k_y^2}{\omega^2/a^2k_z^2-1}\right)\hat\chi=0.
\end{equation}
Note that $\hat\chi$ has a logarithmic singularity at the Alfv\'en resonance $\omega^2=a^2k_z^2$.

A temporal ray packet may be constructed from $\hat\chi$ by Fourier composition,
\begin{equation}
\chi(x,t) = \frac{1}{2\pi} \int_{-\infty}^\infty e^{-\ri\,\omega t}A(\omega)\,\hat\chi(x,\omega)\, \rd\omega   \label{invF}
\end{equation}
where
$A(\omega) =
\sqrt{2 \pi } \,T \exp \left[-
   \half\left(\omega-\omega_0\right)^2T^2
   \right]      $
is a gaussian distribution about $\omega_0$ in the frequency domain. $T$ is the corresponding temporal width of the packet. The harmonic spatial dependence in $\chi$ is implicit: $\chi(x,y,z,t)=\chi(x,t)\exp[\ri(k_yy+k_zz)]$.

The analysis of Paper I, in particular involving Equation (15), shows that $\hat\chi$ depends only on $\sigma$ and $\theta$, and not on frequency $\omega$ except for a trivial translation:
\begin{equation}
\hat\chi(x,\omega)=\check\chi_{\sigma\theta}\left(x-\ln\frac{\omega^2}{\omega_0^2}\right)\, ,    \label{chihat}
\end{equation}
where the reflection point for the central frequency $\omega_0$ has been arbitrarily placed at $x=0$. The resonance is then at $X(\omega)=\ln\sec^2\theta-\ln(\omega^2/\omega_0^2)$, which of course is smeared by the superposition of the continuous range of frequency in the packet. Unlike the spatial ray bundle case (where $X$ was smeared by the range of $\theta$), a single numerical solution $\check\chi$ (for given $\sigma$ and $\theta$) suffices to construct $\chi(x,t)$.

Since the elemental absorption coefficient $\mathscr{A}$ depends only on $\sigma$ and $\theta$ (and thus on $k_y$ and $k_z$), and not on frequency $\omega$, it is clear that the temporal wave packet constructed from waves with a common $\sigma$ and $\theta$ but a range of $\omega$ must have absorption coefficient $\mathscr{A}(\sigma,\theta)$ \emph{precisely}, independent of $\omega_0$ and $T$. We may therefore definitively answer the question of whether absorption may exceed $A_0$ in the negative without further ado. Nevertheless, we illustrate this conclusion with a numerical example.

Having constructed $\chi$, the plasma displacement $\bxi=(\xi_x,\xi_y,0)$ may be formed by integrating the driven Alfv\'en equation
\begin{equation}
\frac{1}{a^2}\pderivd{\bxi}{t}+k_z^2\bxi = \grad_{\!p}\chi\,,
\label{xiDirect}
\end{equation}
where $\grad_{\! p}$ is the $x$--$y$ part of the gradient operator. This clearly shows how $\grad_{\! p}\chi$ acts as a source term for Alfv\'en waves, since $\mathcal{A}=a^{-2}\partial^2/\partial t^2 + k_z^2$ is the free Alfv\'en operator. Equation (\ref{xiDirect}) has the exact solution (assuming $\bxi$ vanishes as $t\to-\infty$)
\begin{equation}
\bxi(x,t) 
 = \frac{a}{k_z}\int_{-\infty}^t \!\! \grad_{\! p}\chi(x,t')\,\sin ak_z(t-t')\,\rd t'  \, . 
 \label{varpar}
\end{equation}

We require expressions for the wave-energy flux [$F_x$] in the $x$-direction and for the wave-energy density [$E$]:
\begin{equation}
F_x = \re\{-\chi^* v_x\}\,,   \label{Fx}
\end{equation}
where $v_x=\partial\xi_x/\partial t$ is the $x$-velocity, and
\begin{equation}
E=\half\left[(|v_x|^2+|v_y|^2)/a^2+k_z^2(|\xi_x|^2+|\xi_y|^2)+|\chi|^2
\right]\, .    \label{E}
\end{equation}
A factor $B^2/\mu$ has been suppressed in both, where $B$ is the background magnetic-field strength and $\mu$ is the magnetic permeability. The three parts of $E$ are respectively the kinetic, the magnetic tension, and the magnetic pressure energies. Assuming that the incident wave enters the system $x>x_L$ (where $x_L<0$) entirely during $t<0$, and leaves during $t>0$, the total energy per unit $y$--$z$ area entering through a left-hand boundary at $x_L$ is $E_+=\int_{-\infty}^0 F_x(x_L,t)\,\rd t>0$, whilst that leaving through the same boundary after reflection is $E_-=\int_0^\infty F_x(x_L,t)\,\rd t<0$. The absorption coefficient is therefore $(E_++E_-)/E_+$, or alternatively $\lim_{t\to\infty}\int_{x_L}^\infty E(x,t)\,\rd x/E_+$. The equivalence of these two expressions is a non-trivial check on the numerics and is attained to good accuracy in our calculations.

\begin{figure}[tbp]
\begin{center}
\includegraphics[width=\textwidth]{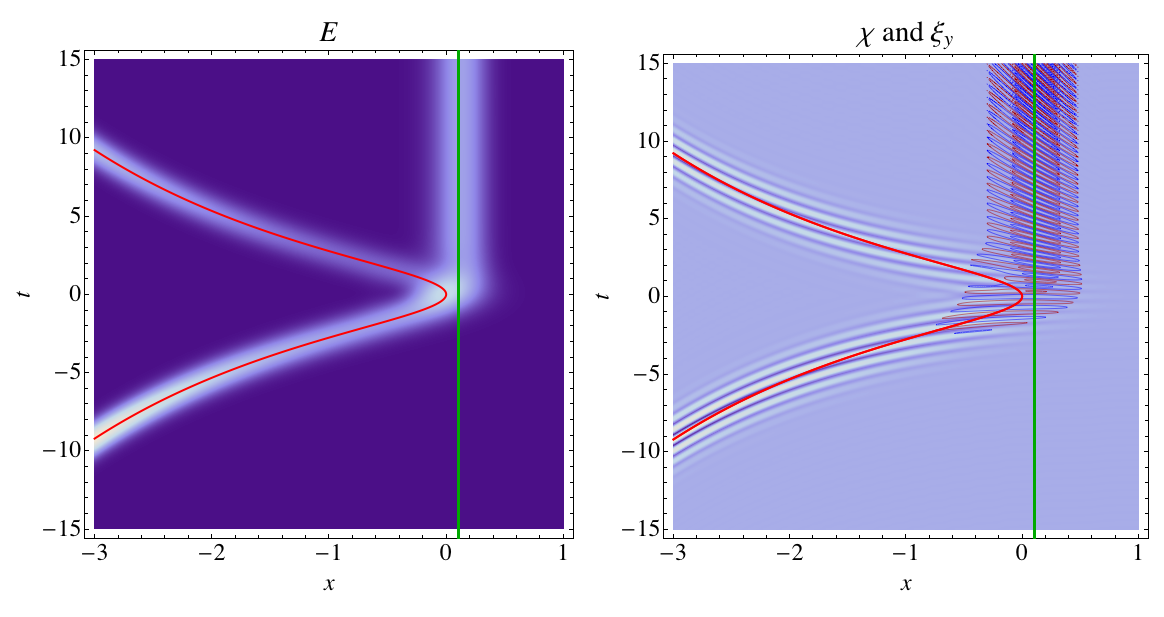}
\caption{Left: Energy density $E$ as a function of $x$ and $t$ for the case discussed in the text. The fast ray bundle enters the box at $x_L=-3$ at around $t=-9.2$ and leaves at about $t=9.2$, leaving 46.4\% of its energy in the resonant layer. The vertical green line is the resonance for the central frequency $\omega_0=9$. Right: $\re\chi$ (background shading) and $\re\xi_y$ (superposed red and blue curves).}
\label{fig:Echixiy}
\end{center}
\end{figure}

The case we examine is $k_y=3$, $k_z=9$ (and hence $\sigma=0.448$, $\theta=18.43^\circ$, $\mathscr{A}=0.4638$, quite close to the maximum possible absorption), $T=1$, with $\omega_0=9$. A total of $2^9\times3=1536$ Fourier modes are employed, and the transform (\ref{invF}) is performed numerically using the fast Fractional Fourier Transform algorithm \cite{bs94}, implemented in parallel. The energy density and displacement vectors are plotted in Figure \ref{fig:Echixiy}. The red curve is the exact ray path predicted by classical ray theory for the central frequency $\omega_0$; \emph{viz.} $\omega_0^2\,t^2=4\kperp^2(e^{-x}-1)$. The left frame shows clearly that the incident fast wave splits into two near the resonance, partially reflecting as a fast wave but also partially remaining on the resonance layer as an Alfv\'en wave. We confirm that $(E_++E_-)/E_+=0.4638$ as expected. The Alfv\'en wave is characterized by its transverse displacement [$\xi_y$] explicitly shown in the right frame, with $\chi$ representing the incident and reflected fast wave, which is compressive. The increasing inclination of these $\xi_y$ wave fronts with time denotes phase mixing.

\begin{figure}[htb]
\begin{center}
\includegraphics[width=\textwidth]{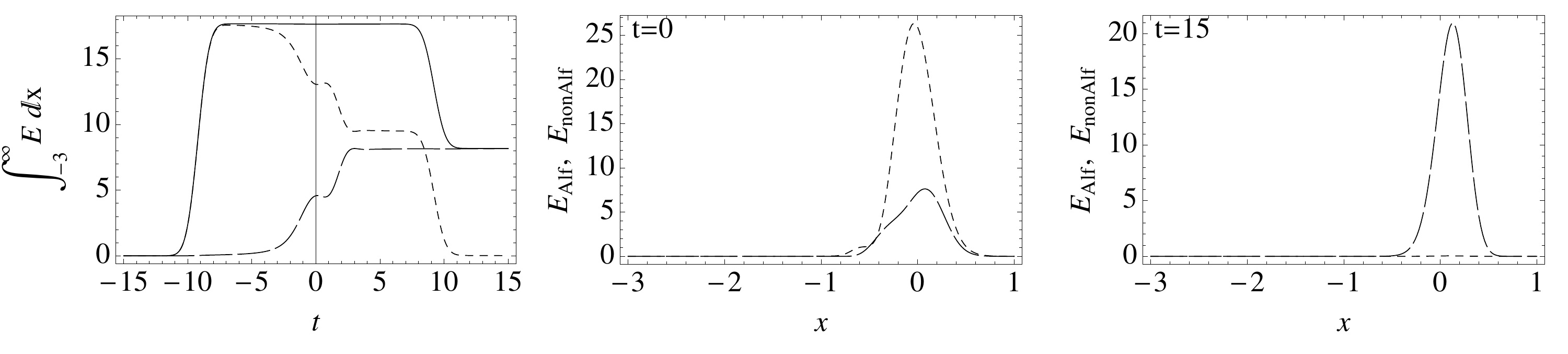}
\caption{Left: Integrated total energy (full curve), Alfv\'enic energy (long dashed) and non-Alfv\'enic energy (short dashed) per unit $y$--$z$ area in the region $-3<x<\infty$ as a function of time $t$. Centre: Alfv\'enic (long dashed) and non-Alfv\'enic (short dashed) energy densities as functions of $x$ at $t=0$. Right: Alfv\'enic and non-Alfv\'enic energy densities at $t=15$.}
\label{fig:Erow}
\end{center}
\end{figure}

Figure \ref{fig:Erow}a displays the total energy per unit area in $x>-3$ showing clearly how it rises monotonically as the wave enters the box from the left, stays constant for a period, then decreases monotonically as the reflected fast wave departs but nevertheless leaves a substantial portion of its energy (the Alfv\'enic part) on the resonance in perpetuity. This, together with Figures \ref{fig:Erow}b and \ref{fig:Erow}c, confirms that the Alfv\'enic energy [$E_\text{Alf}=\half(|v_y|^2/a^2+k_z^2|\xi_y|^2)$] is generated around $t=0$ and is the only significant energy remaining on the resonance layer once the fast wave has departed.

\begin{figure}[tbhp]
\begin{center}
\includegraphics[width=.93\textwidth]{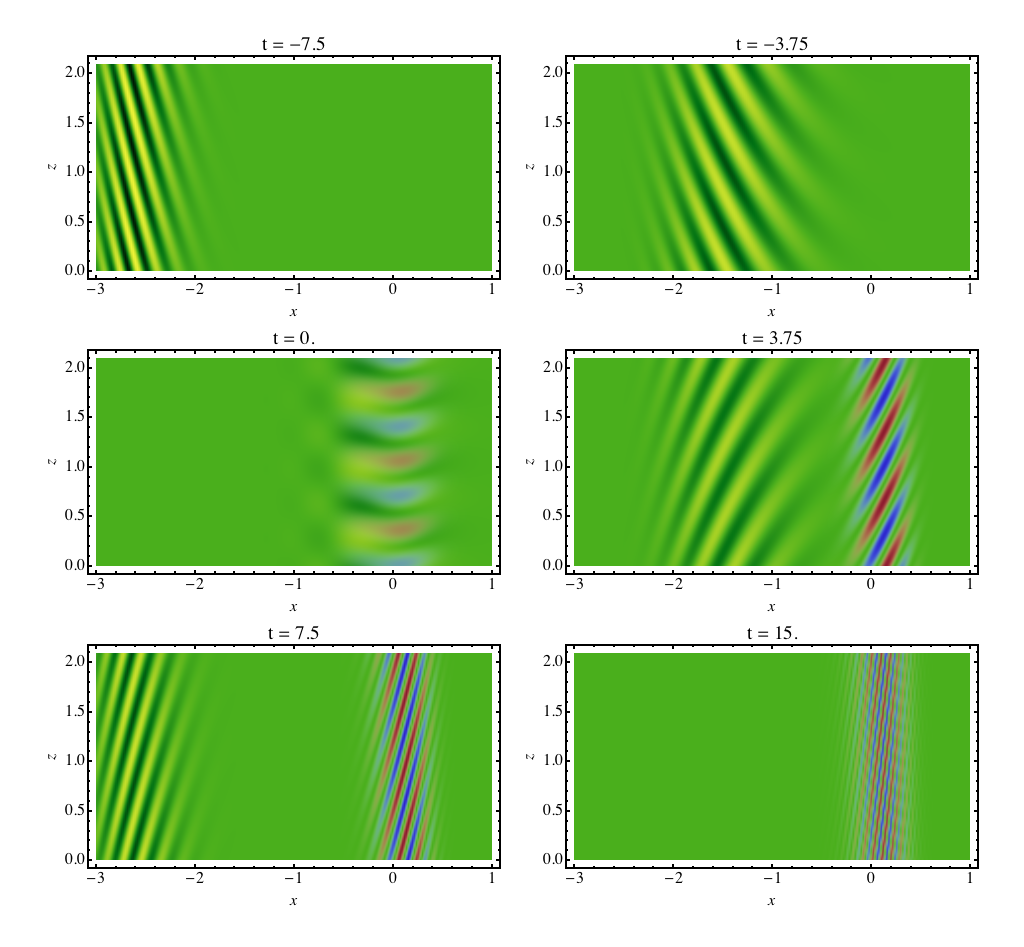}
\caption{Snapshots in time (as labelled) of the ray bundle in $x$-$z$ space, showing both the fast wave ($\chi$, green--yellow shading) and the Alfv\'en wave ($\xi_y$, red--blue). Three wavelengths in $z$ are displayed.}
\label{fig:raygrid}
\end{center}
\end{figure}

Finally, Figure \ref{fig:raygrid} and the accompanying animation 
show the fast wave as a shaded avocado rendering of $\chi$ over-plotted with a red--blue  represention of the Alfve\'n wave in the form of $\xi_y$. They illustrate the reflection and mode conversion very clearly, as well as the phase mixing of the Alfv\'en wave. The movie in particular gives a good physical impression of how the refracting fast wave lines up with the Alfv\'en wave and gives it a ``push along'' in the conversion region. This coincidence of the phase velocities of the converting and converted waves is fundamental to the process.

\section{Conclusions}
We have once again illustrated that constructing a ray packet -- temporal on this occasion -- from singular resonant monochromatic wave solutions provides a clear picture of fast-to-Alfv\'en mode conversion in a cold MHD plasma, with the conversion region localized in both $x$ and $t$. The solutions are smooth everywhere. We find that, since the absorption/conversion coefficient depends only on transverse wave numbers [$k_y$] and [$k_z$], and not on frequency [$\omega$], such a bundle constructed out of waves of different frequency but the same $k_y$ and $k_z$ suffers exactly the absorption predicted by Figure \ref{fig:Asig} and tabulated in Paper I. In particular, it is not possible to engineer a higher conversion coefficient than $A_0=0.4937$. Of course though the frequency is crucial in determining the position of the Alfv\'en resonance.

Although the model adopted here is extremely simple, it is instructive when considering impulsive waves in the solar corona. We may surmise that fast MHD waves generated by flares and other such localized events may easily end up partially mode converting to Alfv\'en waves in resonant layers as they propagate into stronger-field or lower-density regions, and that the width of the pulse should have little influence on the ultimate absorption/conversion fraction, which may not exceed about 50\% when the magnetic field is oriented perpendicular to the direction of inhomogeneity.



\end{document}